\documentclass[12pt]{article}
\pdfoutput=1
\usepackage{subfigure}
\usepackage{amssymb,amsmath}
\usepackage{graphicx}
\usepackage{color}
\usepackage[colorlinks=true
,urlcolor=blue
,citecolor=blue
,linkcolor=blue
,pagecolor=blue
,linktocpage=true
,pdfproducer=medialab
]{hyperref}
\usepackage[a4paper,width=17cm]{geometry}
\makeatletter \renewcommand{\@dotsep}{10000} \makeatother
\usepackage{appendix}

\begin{document}

\begin{center}

 {\Large\bf  Introducing Spin in 2D Quantum Tunneling 
 } \vspace{1cm}

{   Muhammad Adeel Ajaib\footnote{ E-mail: adeel@udel.edu}}

{\baselineskip 20pt \it
Department of Mathematics, Statistics and Physics, Qatar  University,  Doha,  Qatar
 } \vspace{.5cm}

{\baselineskip 20pt \it
   } \vspace{.5cm}

\setcounter{footnote}{0}
\vspace{1.5cm}
\end{center}

\begin{abstract}

We study the quantum tunneling of non-relativistic electrons for two dimensional condensed matter systems. We employ the L{\'e}vy-Leblond equation (which is the analogue of the Dirac equation for non-relativistic fermions) and show that the spin of the particle can be incorporated in the 2D tunneling problem. We derive the transmission and reflection coefficients of spin up and down electrons and show that the sum of these coefficients are consistent with the known results for gapless semiconductors.

\end{abstract}

\newpage

\section{Introduction}\label{intro}

The L{\'e}vy-Leblond equation (LLE) is the analogue of the Dirac equation for non-relativistic fermions. It has been shown that this equation is consistent with fundamental problems in non-relativistic quantum mechanics, for example, the potential step and finite potential barrier problem \cite{LevyLeblond:1967zz, Ajaib:2015uha,Ajaib:2015eer, Ajaib:2017cpl}. It was also shown that this equation is the non-relativistic limit of the Dirac equation and the Pauli Hamiltonian can be obtained from this equation by requiring it to be locally invariant \cite{Ajaib:2015eer}. Furthermore, it was shown that quantized energy level of the Hydrogen atom are obtained when this equation is solved with a Coulomb potential \cite{Ajaib:2017cpl}.

The tunneling of electrons through a potential barrier is one of the foundational problems in quantum mechanics. In this article we discuss the tunneling of non-relativistic electrons through a 2D potential barrier. There have been studies on this topic in recent years in the context of graphene and two dimensional electron gases (2DEG) \cite{Katsnelson:2006, Mohammadpour:2016,Jahani:2013}. The novelty of the analysis we perform herein is that we will introduce spin of the particle in such scattering processes. The analysis performed in this article is applicable to non-chiral electrons in gapless semiconductors \cite{Katsnelson:2006,Esaki:1958}. Recently there has been a growing interests in 2D materials other than graphene \cite{Andres:2016}. The analysis performed in this article can possibly be applied and tested in p-n and n-p-n junctions designed from such materials. 

The paper is organized as follows: In section \ref{sec:lle}, we briefly describe the L{\'e}vy-Leblond equation in (2+1) dimensions and (3+1) dimensions.
In section \ref{sec:2dtunnel}, we solve the problem of 2D tunneling of electrons from a potential barrier by using the LLE in (2+1) dimensions. In section \ref{sec:2dtunnel-4by4}, we study the 2D tunneling problem with 4$\times$4 matrices and solve for the transmission and reflection coefficients of the spin up and down electrons. We conclude in section \ref{conclude}.

\section{The L{\'e}vy-Leblond Equation}\label{sec:lle}

In this section, we briefly describe the LLE and its applications in quantum mechanics (for further details the reader is referred to references \cite{LevyLeblond:1967zz, Ajaib:2015uha,Ajaib:2015eer, Ajaib:2017cpl}). The LLE describes spin half fermions in the non-relativistic limit. Since electrons in condensed matter systems are non-relativistic fermions, the LLE is the more appropriate equation for these systems. 
In 3+1 dimensions the LLE is given by
\begin{eqnarray}
-i \gamma_i \partial_i \psi = (i  \eta \partial_t  + \eta^\dagger m) \psi 
\label{eq:eq-3d}
\end{eqnarray}
where $\gamma_i$ are the Dirac gamma matrices, $\eta$ is a 4$\times$4 nilpotent matrix and we use $\hbar=c=1$ in this section. There are different representations of the $\eta$ matrices, for example, $\eta=(\gamma_0+i \gamma_5)/\sqrt{2}$. In section \ref{sec:diff-rep}, we shall discuss the implications of considering a different representation of these matrices for the potential barrier problem. %There are other representations of the $\eta$ matrices, for example the representation employed in reference \cite{Ajaib:2015uha} is $\eta=i (\gamma_5\gamma_0\gamma_1+\gamma_2)/\sqrt{2}$.

\begin{figure}
\vspace*{-1.2cm}
\centering
\includegraphics[scale=.39]{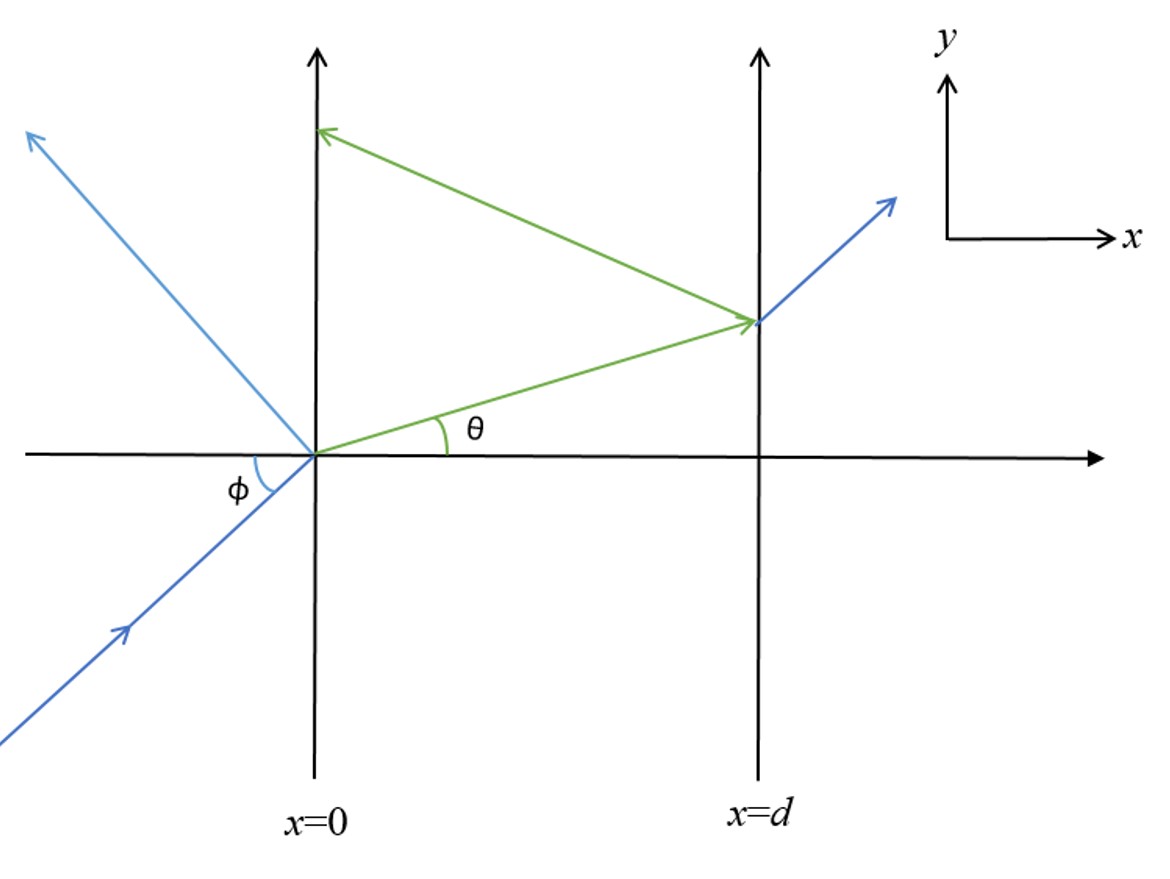}
\includegraphics[scale=.35]{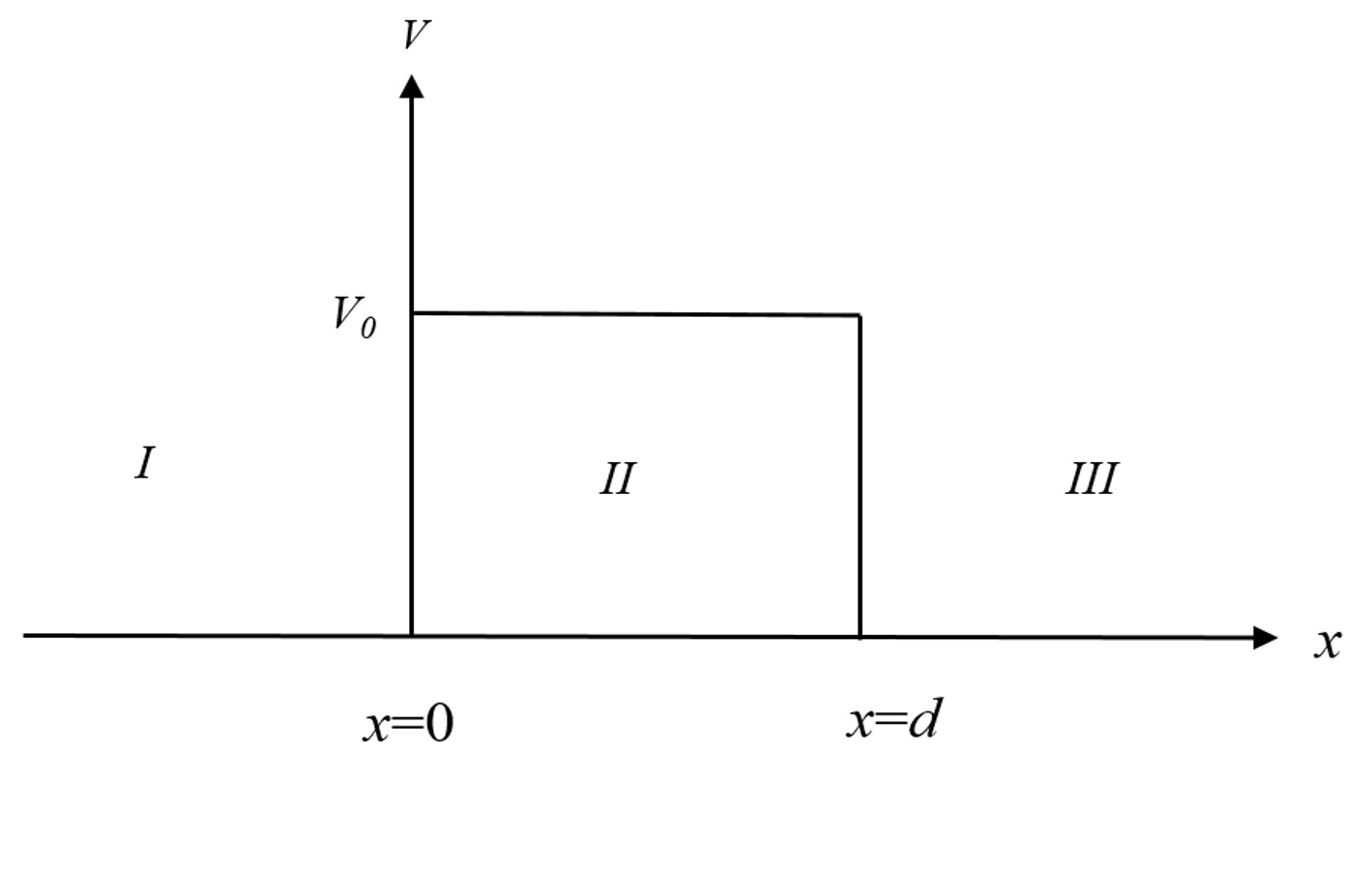}
\caption{Schematic representation of the 2D barrier tunneling problem.}
\label{fig:2dbarrier}
\end{figure}

The (2+1) dimensional version of the L{\'e}vy-Leblond equation for 2$\times$2 matrices, in momentum space, is given by \cite{Ajaib:2015eer}
\begin{eqnarray}
-i \mu_i \partial_i \psi = (i  \eta \partial_t  + \eta^\dagger m) \psi 
%\mu_i p_i   = (  \eta E + \eta^\dagger m)
\label{eq:2dmat2p1}
\end{eqnarray}
where $\mu_1=I$,  $\mu_2=i\sigma_3$ and $\eta= 1/\sqrt{2}(\sigma_1-i \sigma_2)$. It was shown in \cite{Ajaib:2017cpl} that equation (\ref{eq:2dmat2p1}) is the non-relativistic limit of the 2+1 dimensional Dirac equation. In the analysis performed in the following sections we shall consider the LLE in 2D with 2$\times$2  and 4$\times$4 $\eta$ matrices. 

One of the issues in obtaining the Hamiltonian of equation (\ref{eq:eq-3d}) is that the matrix $\eta$ is singular \cite{Ajaib:2017cpl}. In order to obtain the Hamiltonian we replace $\eta \rightarrow \eta^\prime=\eta+\epsilon \eta^\dagger$ and analyze the limit $\epsilon \rightarrow 0$. We thereby obtain the following Hamiltonian for equation (\ref{eq:eq-3d})
\begin{eqnarray}
H=\eta^{\prime -1} (-i \gamma_i \partial_i-m \eta^{\dagger})
\label{eq:hamiltonian}
\end{eqnarray}
where $\eta^\prime=\eta+\epsilon \eta^\dagger$. In the limit $\epsilon \rightarrow 0$, two of the eigenvalues of the Hamiltonian in (\ref{eq:hamiltonian}) correspond to positive finite energy ($E=\vec{p}^2/2m$), whereas the other two energy states correspond to infinite negative energy. The infinite negative energy states were interpreted as the negative sea of filled states \cite{Ajaib:2017cpl} and we discard these in the analysis we perform in this article. Furthermore, the negative energy states in condensed matter systems (holes) can be described by the Hamiltonian, $H=\eta^{\prime -1} (i \gamma_i \partial_i+m \eta^{\dagger})$. In the following sections, we employ equation (\ref{eq:hamiltonian}) in 2D with 2$\times$2  and 4$\times$4 matrices to describe electrons in condensed matter systems and derive the transmission and reflection coefficients for the potential barrier problem. 

It has been shown in references \cite{Ajaib:2015uha,Ajaib:2015eer} that the LLE is consistent with non-relativistic quantum mechanics. The step potential problem and finite potential problem in 1D were discussed in \cite{Ajaib:2015uha,Ajaib:2015eer} and it was shown that the transmission and reflection coefficients of the spin up and down electrons when added, yield the known quantum mechanical results obtained from the Schrodinger equation. We extend these analyses to 2D in this article and show that the results are consistent in this case as well.

\section{Electron Tunneling in 2D with 2$\times$2 Matrices } \label{sec:2dtunnel}

In this section, we analyze the 2D scattering of an electron incident on a potential barrier of height $V_0$ and width $d$ (Figure \ref{fig:2dbarrier}). The (2+1) dimensional version of the L{\'e}vy-Leblond equation for 2$\times$2 matrices (Eq. \ref{eq:2dmat2p1}), in momentum space  ($\psi=u(k_x,k_y) e^{-i k.x}$), is given by
\begin{eqnarray}
%-i \mu_i \partial_i \psi = (i  \eta \partial_t  + \eta^\dagger m) \psi 
 {\it v} 
 \hbar \mu_i k_i \ u(k_x,k_y) = (  \eta E + \eta^\dagger m {\it v}^2)u(k_x,k_y)
\label{eq:2dmat}
\end{eqnarray}
where we have replaced the velocity of light by  the Fermi velocity  $ {\it v}$ of electrons in semiconductors, $ {\it v}\approx 10^6 m/s$. 
The eigenstate of (\ref{eq:2dmat}) corresponding to the dispersion relation $E=\hbar^2\vec{k}^2/2m$ is given by
\begin{eqnarray}
u(k_x,k_y) = 
\left(
\begin{array}{cc}
 \frac{\sqrt{2} m {\it v} }{\hbar(k_x+i k_y)}   \\
 1   \\
\end{array}
\right)
\label{matrix:eta2d}
\end{eqnarray}
The wave function in the three regions shown in Figure \ref{fig:2dbarrier} are given by
\begin{eqnarray}
\psi_{I}(x, y) &=&  u(k_x,k_y) e^{i k_x x} e^{i k_y y}+r \ u(-k_x,k_y) e^{-i k_x x} e^{i k_y y} \\
\psi_{II}(x, y) &=&  a \  u(q_x,k_y) e^{i q_x x} e^{i k_y y} + b \ u(-q_x,k_y) e^{-i q_x x} e^{i k_y y} \\
\psi_{III}(x, y) &=&  t \ u(k_x,k_y) e^{i k_x x} e^{i k_y y}
\end{eqnarray}
where, $k_x=k \cos\phi$, $k_y=k \sin\phi$, $q_x=q \cos\theta$, $q/k=\sqrt{(E-V_0)/E}$ and from conservation of the wave vector in the $y$-direction we have, $k \sin\phi=q \sin\theta$. Applying the continuity of the wavefunction at the boundary $x=0$ and $x=d$ yields the following transmission and reflection coefficients:
\begin{eqnarray}
T_{QM}=|t|^2=\frac{4 k_x^2 q_x^2}{4 k_x^2 q_x^2 \cos^2(d \ q_x) + (k_x^2 + q_x^2)^2 \sin^2(d \  q_x)} \nonumber \\
R_{QM}=|r|^2=\frac{(k_x^2 - q_x^2)^2 \sin^2(d \  q_x)}{4 k_x^2 q_x^2 \cos^2(d \ q_x)  + (k_x^2 + q_x^2)^2 \sin^2(d \  q_x)}
\label{eq:coeff-2by2}
\end{eqnarray}
with $T_{QM}+R_{QM}=1$. This result is consistent with the transmission coefficient obtained by employing the Schrodinger equation. Note that resonance occurs when $q_x d=n\pi$ ($n=0,\pm1,..$) and the barrier is transparent. Moreover, for normal incidence, the transmission coefficient is an oscillating function that varies between 0 and 1. This expression is consistent with the one presented in  \cite{Katsnelson:2006} for non-chiral electrons in gapless semiconductors. In the next section we show that the results for the transmission and reflection coefficients for spin up and down electrons  yields the above expressions.

\begin{figure}
\centering
\includegraphics[scale=.46]{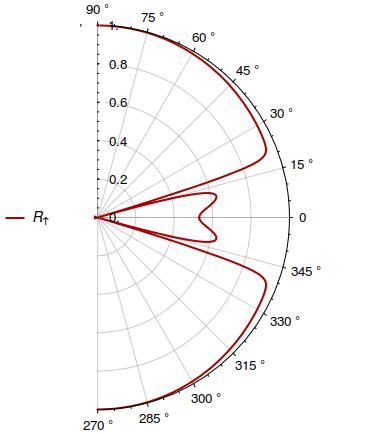}
\includegraphics[scale=.46]{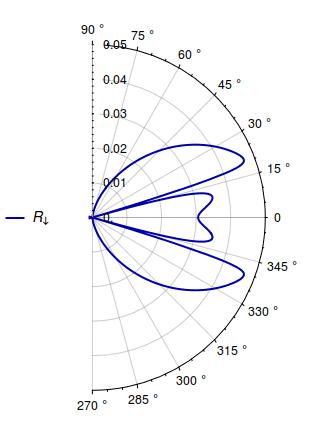}
\caption{The plot shows the reflection coefficients for spin up and down electrons as a function of the incident angle $\phi$, for $E>V_0$, given in equations (\ref{eq:coeffs1}). The red lines show the coefficients of spin up electron and the blue lines correspond to spin down electron. The energy of the incident electrons, the height and width of the potential barrier are chosen to be $E=80$ meV, $V_0=$70 meV and d=10 nm.}
\label{fig:ref-coeff}
\end{figure}

\begin{figure}
\centering
\includegraphics[scale=.47]{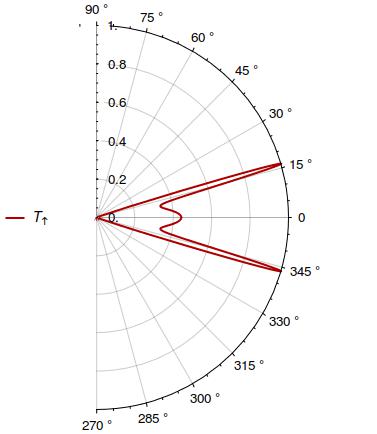}
\caption{The plot shows the transmission coefficient for spin up electrons as a function of the incident angle $\phi$, for $E>V_0$, given in equations (\ref{eq:coeffs1}). The red lines show the coefficients of spin up electron and the blue lines correspond to spin down electron. The energy of the incident electrons, the height and width of the potential barrier are chosen to be $E=80$ meV, $V_0=$70 meV and d=10 nm.}
\label{fig:tran-coeff}
\end{figure}

\section{Electron Tunneling in 2D with 4$\times$4 Matrices } \label{sec:2dtunnel-4by4}
\begin{figure}
\centering
\includegraphics[scale=.6]{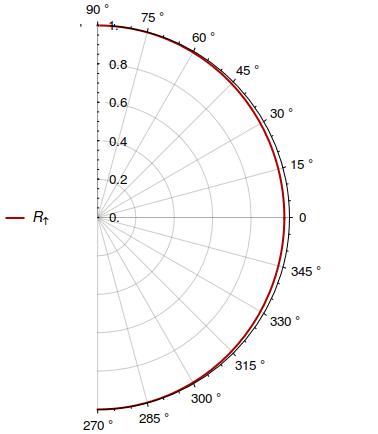}
\includegraphics[scale=.62]{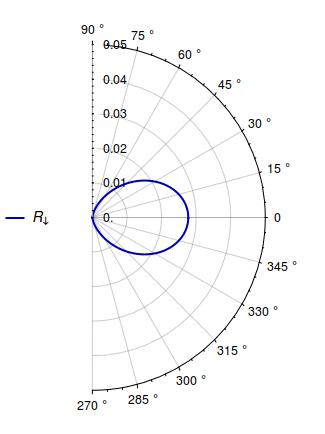}
\caption{The plot shows the reflection coefficients for spin up and down electrons as a function of the incident angle $\phi$, for $E<V_0$ given in equations (\ref{eq:rfcoeffs-elthv0}). The red lines show the coefficients of spin up electron and the blue lines correspond to spin down electron. The energy of the incident electrons, the height and width of the potential barrier are chosen to be $E=40$ meV, $V_0=$50 meV and d=10 nm.}
\label{fig:ref-coeff-eltv0}
\end{figure}

In this section, we demonstrate how the LLE can be employed to incorporate the spin of the particle in the scattering process discussed in the previous section  (Figure \ref{fig:2dbarrier}). We analyze the case of a spin up electron with energy $E$ incident on a potential barrier of height $V_0$ and width $d$.  
In momentum space, equation (\ref{eq:eq-3d}) is given by
\begin{eqnarray}
 \hbar v \gamma_i k_i \ u(k_x,k_y) = (\eta E  + \eta^\dagger m v^2) u(k_x,k_y)
\end{eqnarray}
where $\gamma_i=(\gamma_1,\gamma_2)$. The spin up and down eigenstates corresponding to the dispersion relation $E=\hbar^2\vec{k}^2/2m$ are given by
\begin{eqnarray}
u(k_x,k_y) = 
\left(
\begin{array}{cc}
 1   \\
 0   \\
 -i + \frac{4 i m^2 v^2}{2 m^2 + k_x^2 + k_y^2}\\
 \frac{2 \sqrt{2} m v (k_x - i k_y)}{2 m^2 v^2 + k_x^2 + k_y^2}
\end{array}
\right)
\label{matrix:eta2d}
\end{eqnarray}

\begin{eqnarray}
v(k_x,k_y) = 
\left(
\begin{array}{cc}
 0   \\
 1   \\
 \frac{2 \sqrt{2} m v (k_x - i k_y)}{2 m^2 v^2 + k_x^2 + k_y^2}\\
  -i + \frac{4 i m^2 v^2}{2 m^2 v^2 + k_x^2 + k_y^2}
\end{array}
\right)
\label{matrix:eta2d}
\end{eqnarray}
The wave functions in the three regions shown in Figure \ref{fig:2dbarrier} are given by
\begin{eqnarray}
\psi_{I}(x, y) &=&  (u(k_x,k_y) e^{i k_x x} +r_1 \ u(-k_x,k_y) e^{-i k_x x} +r_2 \ v(-k_x,k_y) e^{-i k_x x}) e^{i k_y y} \nonumber \\ 
\psi_{II}(x, y) &=&   (a_1   u(q_x,k_y) e^{i q_x x} + a_2   u(q_x,k_y) e^{i q_x x} + b_1  u(-q_x,k_y) e^{-i q_x x} +b_2  u(-q_x,k_y) e^{-i q_x x} ) e^{i k_y y} \nonumber \\
\psi_{III}(x, y) &=&  (t_1 \ u(k_x,k_y) e^{i k_x x}  + t_2 \ v(k_x,k_y) e^{i k_x x} ) e^{i k_y y} \nonumber
\end{eqnarray}
Applying the continuity of the wavefunction at the boundary $x=0$ and $x=d$, yields the following transmission and reflection coefficients:
\begin{eqnarray}
T_1 &=& |t_1|^2 = \frac{4 k_x^2 q_x^2}{4 k_x^2 q_x^2 \cos^2(d \ q_x) + (k_x^2 + q_x^2)^2 \sin^2(d \  q_x)}\nonumber  \\
T_2 &=& |t_2|^2=0 \nonumber  \\
R_1 &=& |r_1|^2=\frac{(4 m^4 v^4 + 4 m^2 v^2 (-k_x^2 + k_y^2) + (k_x^2 + k_y^2)^2) (k_x - q_x)^2 (k_x + 
   q_x)^2 \sin(d q_x)^2}{(2 m^2 v^2 + k_x^2 + 
   k_y^2)^2 (4 k_x^2 q_x^2 \cos(d q_x)^2 + (k_x^2 + q_x^2)^2 \sin(d q_x)^2)} \nonumber \\
R_2 &=& |r_2|^2=\frac{8 m^2 v^2 k_x^2 (k_x^2 - q_x^2)^2 \sin(d q_x)^2}{(2 m^2 v^2 + k_x^2 + 
   k_y^2)^2 (4 k_x^2 q_x^2 \cos(d q_x)^2 + (k_x^2 + q_x^2)^2 \sin(d q_x)^2)}
\label{eq:coeffs1}
\end{eqnarray}
where
\begin{eqnarray}
T_1+T_2+R_1+R_2=1
\end{eqnarray}
and $R_1+R_2=R_{QM}$. In Figures (\ref{fig:ref-coeff}) and (\ref{fig:tran-coeff}) we display the plots of the reflection and transmission coefficients for spin up and down electrons. For these plots, the height of the potential barrier is chosen to be $V_0=$70 meV and the width as $d$=10 nm. The red lines show the coefficients of spin up electron and the blue lines correspond to spin down electron. The transmitted electron is always spin up and probability of the transmitted electron to be spin down is zero ($T_2=0$). The reflected electron is most likely a spin up electron and there is a small probability ($< 5 \%)$ that the reflected electron will be spin down. For $E<V_0$, the transmission coefficient $T \simeq 0$ and the reflection coefficients are shown in Figure \ref{fig:ref-coeff-eltv0}. The reflected electron is most likely spin up with a small probability ($<3 \%$) that it will flip its spin.

\begin{figure}
\centering
\includegraphics[scale=.6]{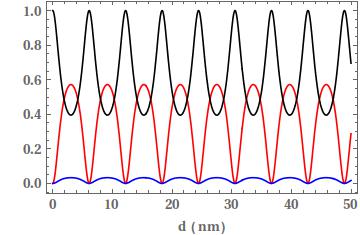}
\includegraphics[scale=.6]{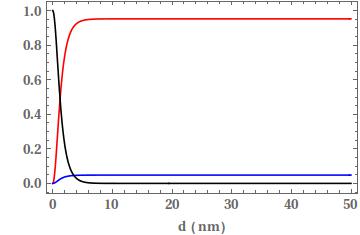}
\caption{Transmission and reflection coefficients (\ref{eq:coeffs1}) for normally incident electrons ($\phi=0$) as a function of the width $d$ of the tunnel barrier for $E>V_0$ (left) and $E<V_0$ (right). The red and blue lines show the reflection coefficients of spin up electron and spin down electron whereas the black represents the transmission coefficient of spin up electron. The width of the barrier is chosen to be $d$=10 nm for both the plots whereas we choose $E=80$ meV, $V_0=$70 meV (left) and $E=70$ meV, $V_0=$80 meV (right).}
\label{fig:all-coeffs-norm}
\end{figure}

Figure \ref{fig:all-coeffs-norm} shows the coefficients for normally incident ($\phi =0$) electrons as a function of barrier width $d$, for $E>V_0$ (left) and $E<V_0$ (right). For  $E>V_0$, the reflection (red for spin up and blue for spin down) and transmission coefficients (black) are oscillating functions of the barrier width $d$. For $E<V_0$, as the width of the barrier increases the probability of transmission decreases significantly. We can see that for this case there is a small probability $\sim 5 \%$ that the electron will flip its spin upon reflection.

\begin{figure}
\vspace*{-1.2cm}
\centering
\includegraphics[scale=.55]{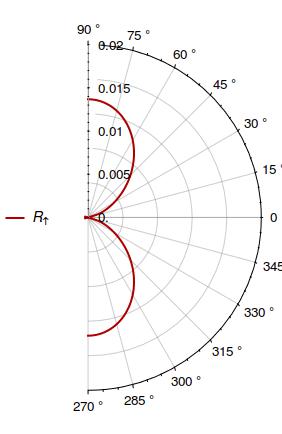}
\includegraphics[scale=.55]{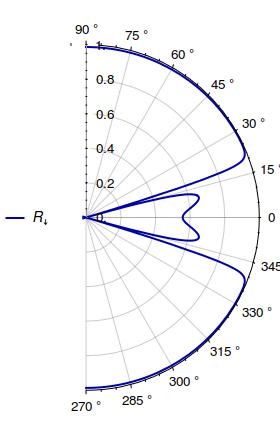}
\caption{The plot shows the reflection coefficients for spin up and down electrons for $E>V_0$ given in equations (\ref{eq:rfcoeffs-elthv0}). The red lines show the coefficients of spin up electron and the blue lines correspond to spin down electron. The energy of the incident electrons, the height and width of the potential barrier are chosen to be $E=80$ meV, $V_0=$70 meV and d=10 nm.}
\label{fig:ref-coeff2-egtv0}
\end{figure}

\subsection{A Different Representation of Matrices } \label{sec:diff-rep}

We will now show that a different representation of the $\eta$ matrices can lead to different implications for the scattering coefficients of a particle in 2D scattering. In this section we choose $\eta=-i (\gamma_2+\gamma_5)/\sqrt{2}$ in equation (\ref{eq:hamiltonian}). This representation predicts the same transmission coefficients but different reflection coefficients for spin up and down electrons in the 2D finite barrier problem. Note however that the sum of the transmission and reflection coefficients for the spin up and down is always equal to the ones presented in equations (\ref{eq:coeff-2by2}). 
This representation predicts a large probability for the electron to flip its spin upon reflection. The transmission coefficients are still the same as before whereas the reflection coefficients for spin up and down electrons are now given by
\begin{eqnarray}
R_1 &=&\frac{m^2 v^2 }{k_y^2+m^2 v^2 } |r_1|^2=\frac{1 }{k_y^2+m^2 v^2 }\frac{k_y^2 (k_x^2 - q_x^2)^2}{2  ((k_x^2 + q_x^2)^2 + 4 k_x^2 q_x^2 \cot(d \ q_x)^2)} \\
R_2 &=&\frac{m^2 v^2 }{k_y^2+m^2 v^2 } |r_2|^2=\frac{1}{k_y^2+m^2 v^2 }\frac{(k_x^2 - q_x^2)^2 (k_y^2 + 2 m^2 v_F^2)}{2 ((k_x^2 + q_x^2)^2 + 4 k_x^2 q_x^2 \cot(d \ q_x)^2)}
\label{eq:rfcoeffs-elthv0}
\end{eqnarray}
where
\begin{eqnarray}
R_1+R_2=R_{QM}
\end{eqnarray}
Note also that there are finite number of representations for the $\eta$ matrices and experimental observations are needed to test the predictions of this analysis. The plots of these coefficients are display in Figure \ref{fig:ref-coeff2-egtv0}. The energy of the incident electrons, the height and width of the potential barrier are chosen to be $E=80$ meV, $V_0=$70 meV and d=10 nm. We can see that the conclusions from these graphs are essentially opposite to those from the previous case. The probability of electron to be reflected as spin flipped is significant whereas the probability that it retains its spin orientation is very small ($\lesssim 1\%$).

\section{Conclusion} \label{conclude}

We have analyzed the 2D scattering of electrons from a finite potential barrier. We employed the L{\'e}vy-Leblond equation and showed that the spin of the electron can be incorporated in this problem. We studied the problem of a spin up electron incident on the barrier and showed that the sum of the reflection and transmission coefficients for spin up and down electrons yields results obtained from the Schrodinger equation which are known in literature. These results can possibly be tested for non-chiral electrons in 2 dimensional electron gases and gapless semiconductors.

\section{Acknowledgments}
The author would like to thank Ahsan Zeb and Fariha Nasir for useful discussions and suggestions.


\begin{thebibliography}{99}

%\cite{LevyLeblond:1967zz}
\bibitem{LevyLeblond:1967zz} 
  J.~M.~L{\'e}vy-Leblond,
  %``Nonrelativistic particles and wave equations,''
  Commun.\ Math.\ Phys.\  {\bf 6}, 286 (1967).
  doi:10.1007/BF01646020
  %%CITATION = doi:10.1007/BF01646020;%%
  %146 citations counted in INSPIRE as of 26 Jul 2016

%\cite{Ajaib:2015uha}
\bibitem{Ajaib:2015uha} 
  M.~A.~Ajaib,
  %``A Fundamental Form of the Schr{\"o}dinger Equation,''
  Found.\ Phys.\  {\bf 45}, no. 12, 1586 (2015)
  doi:10.1007/s10701-015-9944-z
  [arXiv:1502.04274 [quant-ph]].
  %%CITATION = doi:10.1007/s10701-015-9944-z;%%

%\cite{Ajaib:2015eer}
\bibitem{Ajaib:2015eer} 
  M.~A.~Ajaib,
  %``Non-Relativistic Limit of the Dirac Equation,''
  arXiv:1511.07901 [quant-ph].
  %%CITATION = ARXIV:1511.07901;%%

%\cite{Ajaib:2017cpl}
\bibitem{Ajaib:2017cpl} 
  M.~A.~Ajaib,
  %``The Hydrogen Atom and the Equivalent Form of Levy-Leblond Equation''
  Chin. Phys. Lett. Vol. 34, No. 5 (2017) 050301. [arXiv:1511.07901 [quant-ph]].
  %%CITATION = ARXIV:1511.07901;%%

%\cite{Katsnelson:2006}
\bibitem{Katsnelson:2006}
M. I. Katsnelson, K. S. Novoselov, and A. K. Geim, Nat. Phys. 2, 620 (2006).

%\cite{Mohammadpour:2016}
\bibitem{Mohammadpour:2016}
H. Mohammadpour, ACTA PHYSICA POLONICA A, Vol. 130 (2016).

%\cite{Jahani:2013}
\bibitem{Jahani:2013}
Dariush Jahani (2013). Electronic Tunneling in Graphene, New Progress on Graphene Research, Prof. Jian Ru Gong (Ed.), InTech, DOI: 10.5772/51980. %Available from: https://www.intechopen.com/books/new-progress-on-graphene-research/electronic-tunneling-in-graphene

%\cite{Andres:2016}
\bibitem{Andres:2016}
Andres Castellanos-Gomez, Nature Photonics 10, 202–204 (2016)

%\cite{Esaki:1958}
\bibitem{Esaki:1958}
Esaki, L.  Phys. Rev. 109, 603–604 (1958).


  
\end{thebibliography}
\end{document}